\documentclass[showpacs,superscriptaddress,twocolumn,amsmath,aps]{revtex4}
\usepackage{graphicx,color}
\usepackage{bm}
\usepackage[hypertex]{hyperref}

\newcommand{\be}{\begin{equation}}
\newcommand{\ee}{\end{equation}}
\newcommand{\bea}{\begin{eqnarray}}
\newcommand{\eea}{\end{eqnarray}}
\newcommand{\bsube}{\begin{subequations}}
\newcommand{\esube}{\end{subequations}}

\newcommand{\Eq}[1]{Eq.\,(\ref{#1})}

\newcommand{\la}{\langle}
\newcommand{\ra}{\rangle}

\usepackage{amsfonts}
\usepackage{amsmath}
\usepackage{amssymb}
\usepackage{graphicx}
\usepackage{type1cm}        
\usepackage{times}          
\usepackage{bm}
\usepackage{color}                      

\newcommand{\bet}{\beta}
\newcommand{\gam}{\gamma}

\newcommand{\eps}{\epsilon}
\newcommand{\vep}{\varepsilon}

\newcommand{\lam}{\lambda}

\newcommand{\Del}{\Delta}

\newcommand{\Ome}{\Omega}

\newcommand{\beq}{\begin{equation}}
\newcommand{\eeq}{\end{equation}}
\newcommand{\beqn}{\begin{eqnarray}}
\newcommand{\eeqn}{\end{eqnarray}}
\newcommand{\bsub}{\begin{subequations}}
\newcommand{\esub}{\end{subequations}}

\newcommand{\re}{\nonumber\\}
\newcommand{\ket}[1]{{\left| #1 \right\rangle }}
\newcommand{\bra}[1]{{\left\langle #1 \right|}}

\newcommand{\LL}{\mathcal {L}}
\newcommand{\RR}{\mathcal {R}}

\newcommand{\cdg}{c^\dagger}
\newcommand{\ddg}{d^\dagger}
\newcommand{\fdg}{f^\dagger}
\newcommand{\wdg}{w^\dagger}

\begin{document}


\title{ Demonstrating nonlocality induced teleportation
through Majorana bound states in a semiconductor nanowire }

\author{Peiyue Wang}
\affiliation{Department of Physics, Beijing Normal University,
Beijing 100875, China}

\author{Yunshan Cao }
\affiliation{School of Physics, Peking University, Beijing 100871,China}

\author{Ming Gong}
\affiliation{Department of Physics and Astronomy, Washington State University,
Pullman, WA 99164 USA}

\author{Shu-Shen Li}
\affiliation{State Key Laboratory for Superlattices and Microstructures,
Institute of Semiconductors, Chinese Academy of Sciences,
P.O. Box 912, Beijing 100083, China}

\author{Xin-Qi Li}
\email{lixinqi@bnu.edu.cn}
\affiliation{Department of Physics, Beijing Normal University,
Beijing 100875, China}

\date{\today}

\begin{abstract}
It was predicted by Tewari {\it et al}
[Phys. Rev. Lett. {\bf 100}, 027001 (2008)]
that a {\it teleportationlike} electron transfer phenomenon
is one of the novel consequences of the existence of
Majorana fermion, because of the inherently nonlocal nature.
In this work we consider a concrete realization and measurement
scheme for this interesting behavior, based on a setup consisting of
a pair of quantum dots which are tunnel-coupled to a semiconductor
nanowire and are jointly measured by two point-contact detectors.
We analyze the teleportation dynamics in the presence of measurement
backaction and discuss how the teleportation events can be identified
from the current trajectories of strong response detectors.
\end{abstract}

\pacs{73.21.-b,74.78.Na,73.63.-b,03.67.Lx}

\maketitle

\graphicspath{{figure/}}

The search for Majorana fermions in solid states
has been attracting a great deal of attention in the past years
\cite{Kit01,Fu08,Ore10,Zhang,Sau10,Lut10,Ali10,Sau12}.
In solid states, it has been predicted that the Majorana
bound states (MBSs) can appear for instance in the
5/2 fractional quantum Hall system\cite{Moore1991}
and the $p$-wave superconductor and superfluid \cite{Read2000}.
In particular, an effective $p$-wave superconductor
can be realized by a semiconductor nanowire
with Rashba spin-orbit interaction and Zeeman splitting and
in proximity to an $s$-wave superconductor \cite{Ore10,Zhang,Sau10,Sau12}.
This opens a new avenue of searching for Majorana fermions
using the most conventional materials.
Also, some demonstrating schemes were proposed based on various transport
signatures, including the tunneling spectroscopy
which may reveal characteristic zero-bias conductance peak \cite{DS01,Liu11}
and peculiar noise behaviors \cite{Bol07,Law09},
the nonlocality nature of the MBSs \cite{Nil08,LeeDH08},
and the 4$\pi$ periodic Majorana-Josephson currents
\cite{Kit01,Lut10,Ore10,Fu09}.
In the aspect of experiment, exotic signatures that may
reveal the existence of MBSs have been observed
in the system of semiconductor nanowire in proximity
to an $s$-wave superconductor\cite{Kou12,MF2,MF3,MF4}.

An inevitable consequence of the existence of Majorana zero modes is
that the fermion quasiparticle excitations are inherently nonlocal.
To be specific, let us consider a semiconductor nanowire in the
topological regime which thus supports the MBSs at the two ends
\cite{Ore10,Sau10,Sau12,Kou12},
and denote the MBSs by Majorana operators $\gamma_1$ and $\gamma_2$.
They are related to the regular fermion operator in
terms of $f^{\dagger}=(\gamma_1+i\gamma_2)/\sqrt{2}$
and its Hermitian conjugate $f$.
This connection implies some remarkable consequences.
For instance, if an electron with energy smaller than
the energy gap between the Majorana zero mode
and other exited states is injected into the system,
we can only have the excitation described by $f$ and $f^{\dagger}$.
This means that a single electron is ``split" into
two Majorana bound states which are, however, spatially separated.
In this work, instead of exploiting certain {\it indirect}
transport signatures, we discuss a possible and very {\it direct}
way to demonstrate this intrinsic {\it nonlocality} of
the paired Majorana modes.

\begin{figure}[!htbp]
  \centering
  \includegraphics[width=7cm]{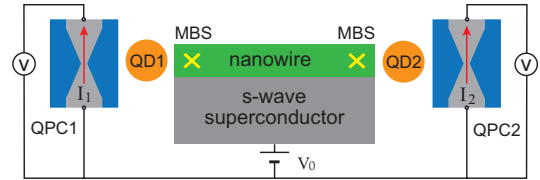}\\
  \caption{
Schematic setup of using two point-contact detectors to demonstrate
the Majorana-nonlocality-induced {\it teleportationlike}
electron transfer between two remote quantum dots.
The semiconductor nanowire is in contact with an $s$-wave
superconductor, so that under appropriate conditions
a pair of Majorana bound states (MBS) are anticipated
to appear at the ends of the nanowire.
Here we show the schematic closed circuit, in which
the chemical potential of the superconductor and the
bias voltages across the detectors are explicitly defined.
}\label{FIG1}
\end{figure}

The proposed scheme is schematically displayed in Fig.\ 1,
where the two MBSs, generated at the ends of the nanowire,
are tunnel-coupled to two quantum dots (QDs), respectively.
Moreover, the QDs are jointly probed by the nearby
quantum-point-contact (QPC) detectors.
This proposal is motivated by
the nowadays state-of-the-art technique,
which enables the QPC current to sensitively probe
an extra single electron in the nearby quantum dot \cite{Ens06}.
In Ref.\ \cite{LeeDH08}, an equivalent ``dot-MBSs-dot" system
is analyzed by assuming an extra electron
initially in one of the QDS and considering
its transmission through the MBSs in a vanished hybridization limit.
Corresponding to the nanowire realization in Fig.\ 1,
their prediction indicates that, in a ``long-wire" limit,
the electron can transmit through
the nanowire on a finite (short) timescale,
revealing thus a ``teleportation" or ``superluminal" phenomenon.
In our present work, following Ref.\ \cite{LeeDH08},
we call this {\it ultrafast} transfer behavior {\it teleportation},
which is actually a remarkable consequence of Majorana's nonlocality.
Related to the scheme of joint-measurements shown in Fig.\ 1,
we will carry out the teleportation dynamics
under the influence of measurement backaction,
and discuss how the teleportation events can be identified
from the current trajectories of strong response detectors.
Also, we will present an interpretation/understanding
to the ``teleportation"  (``superluminal") issue.

{\it Model.}---
The setup of Fig.\ 1 can be described
by the following Hamiltonian
\beq\label{H-t}
    H = H_{sys}+ H_{pc}.
\eeq
The {\it system} Hamiltonian, $H_{sys}$, describes the MBSs
plus the single-level QDs and their tunnel coupling as follows
\cite{Lut10,Liu11,Bol07,Law09,LeeDH08}
\beq\label{Ham1}
    H_{sys}=i\frac{\eps_M}{2}\gam_1\gam_2
    +\sum_{j=1,2}\big[ \eps_j\ddg_j d_j+\lam_j(\ddg_j-d_j)\gam_j \big].
\eeq
Here $\gam_1$ and $\gam_2$ are the Majorana operators
associated with the two MBSs at the ends of the nanowire.
The two MBSs interact with each other by a strength $\eps_M\sim e^{-L/\xi}$,
which damps exponentially with the length ($L$) of the nanowire,
with a characteristic length of the superconducting coherent length ($\xi$).
$d_1(\ddg_1)$ and $d_2(\ddg_2)$ are the annihilation (creation)
operators of the two single-level quantum dots,
while $\lam_1$ and $\lam_2$ are their coupling amplitudes to the MBSs.
In practice, it will be convenient to switch from the Majorana representation
to the regular fermion one, through the transformation of
$\gam_1=i(f-\fdg)$ and $\gam_2=f+\fdg$.
We can easily check that $f$ and $f^{\dagger}$ satisfy
the anti-commutative relation, $\{f,\fdg\}=1$.
After an additional local gauge transformation,
$d_1\to id_1$, we reexpress Eq.\ (\ref{Ham1}) as
\beqn
    H_{sys}&=&\eps_M (\fdg f-\frac{1}{2})
    +\sum_{j=1,2}\big[ \eps_j\ddg_j d_j+\lam_j(\ddg_j f+\fdg d_j)\big]  \re
    &&-\lam_1(\ddg_1\fdg+f d_1)+\lam_2(\ddg_2\fdg+f d_2).
\eeqn
It should be noticed that the tunneling terms in
this Hamiltonian only conserve charge modulo $2e$.
This reflects the fact that a pair of electrons
can be extracted out from the superconductor condensate
and can be absorbed by the condensate.

The other Hamiltonian in \Eq{H-t}, $H_{pc}$, is for the two
point-contacts which reads
\beqn
    H_{pc}=\sum_{j=1,2}\sum_{l_j,r_j}\big[ \big(\vep_{l_j}\cdg_{l_j} c_{l_j}
    +\vep_{r_j}\cdg_{r_j} c_{r_j} \big)     \re
    + \big( w_{j}\cdg_{l_j} c_{r_j}+ {\rm H.c.} \big) \big].
\eeqn
This Hamiltonian simply describes electron tunneling
through a potential barrier between two electronic reservoirs
(with electron creation and annihilation operators,
$\cdg_{l_j(r_j)}$ and $c_{l_j(r_j)}$).
We assume that the tunneling amplitudes ($w_j$) are approximately
of energy independence. Thus $w_j$ does not depend on the
associated states ``$l_j$" and ``$r_j$".
However, in $w_j$ we should include the effect of the nearby quantum dot,
since its occupation would change the tunneling amplitudes.
We account for this effect in terms of $w_j=\Ome_j+\Del\Ome_j\ddg_j d_j$.

{\it Teleportation.}---
Let us consider the transfer problem of an {\it extra} electron
between the two quantum dots, which is assumed
initially in the left quantum dot.

In this part we assume a simpler setup in the absence of
the point-contact detectors \cite{LeeDH08}.

In particular, we consider the weak interaction limit $\eps_M\to 0$,
in order to reveal the remarkable {\it teleportation} behavior.
Using the transformed representation, $\ket{n_1, n_M, n_2}$ describes
the possible charge configuration of the dot-MBSs-dot system,
where $n_{1(2)}$ and $n_M$ denote, respectively, the electron number
(``0" or ``1") in the left (right) dot and the central MBSs.
Totally, we have eight basis states, which can be
divided into two subspaces: $\ket{100}, \ket{010},\ket{001},\ket{111}$
with odd parity (electron numbers);
and $\ket{110},\ket{101},\ket{011},\ket{000}$ with even parity.
Associated with our specific initial condition, we will
only have the odd-parity states involved in the state evolution.
Moreover, for simplicity, we assume $\lam_1=\lam_2=\lam$
and $\eps_1=\eps_2=0$ throughout this work.

Simple calculation can give the occupation probabilities
of the left and right dots, respectively, as
$P_1(t)=\cos^2(\lam t)$ and $P_2(t)=\sin^2(\lam t)$.
Here, for each of the probabilities,
it contains two possible occupations:
$\ket{100}$ and $\ket{111}$ for $P_1(t)$;
$\ket{001}$ and $\ket{111}$ for $P_2(t)$.
Now, we introduce (extract) the partial probability
$P^{(1)}_2(t)=|\la 001 | e^{-iH_{sys}t}\ket{100}|^2$ from $P_2(t)$,
which has also a simple form, $P^{(1)}_2(t)=\sin^4(\lam t)$.
Similarly, we may define
$P^{(2)}_2(t)=|\la 111 | e^{-iH_{sys}t}\ket{100}|^2$,
which can be obtained simply by
$P^{(2)}_2(t)=P_2(t)-P^{(1)}_2(t)=\sin^2(\lam t)\cos^2(\lam t)$.
Based on these simple manipulations, of great interest is
the result of $P^{(1)}_2(t)$, since it implies that,
even in the limit of $\eps_M\to 0$ (very ``long" nanowire),
the electron in the left dot can transmit through the MBSs
and appear in the right dot on some finite (short) timescale.
This is the remarkable ``teleportation" phenomenon
discussed in Ref.\ \cite{LeeDH08}
which, surprisingly, holds a ``superluminal" feature.
In the following, to prove this teleportation behavior,
we propose to use QPC detectors to perform
a {\it coincident} measurement of both the occupation numbers
of the left and the right dots.
This type of measurement can distinguish
the process responsible for $P^{(1)}_2(t)$
from that responsible for $P^{(2)}_2(t)$.

{\it Demonstration.}---
Now we turn to the measurement setup of Fig.\ 1.
Physically, the measurements will cause backaction on the
charge transfer dynamics in the central dot-MBSs-dot system.
This effect can be described by a master equation,
formally expressed as \cite{Li0405}
\beq\label{ME-1}
    \dot{\rho}=-i\LL\rho-\RR\rho .
\eeq
The first term denotes $\LL\rho=[H_{sys},\rho]$,
and the second term describes the measurement backaction.
More specifically,
$\RR\rho=\frac{1}{2} \sum_{j=1,2}
\{ [\wdg_j, \tilde{w}_j^{(-)}\rho-\rho\tilde{w}_j^{(+)}]+ {\rm H.c.} \}$,
where $\tilde{w}_j^{(\pm)}=C_j^{(\pm)}(\pm\LL)w_j$.
$C_j^{(\pm)}(\pm\LL)$ are the Liouvillian counterparts of
the QPC spectral functions $C_j^{(\pm)}(\pm\omega)$,
which were obtained explicitly in Ref.\ \cite{Li0405}.
In this work, we restrict to a wideband limit
and large bias condition for the point-contact detectors,
which allow us to approximate $C_j^{(\pm)}(\pm\LL)$ by $C_j^{(\pm)}(0)$.
More explicitly, we have \cite{Li0405}:
$C_j^{(\pm)}(0)=\pm 2\pi g_L g_R eV_j/(1-e^{\mp \bet e V_j})$,
where $g_{L(R)}$ is the density-of-states of the QPC reservoir,
$V_j$ is the applied voltage, and $\beta$ is the inverse temperature.
Under these considerations, \Eq{ME-1} becomes
the Lindblad-type master equation, with
$\RR \rho = - \sum_{j=1,2} \Gamma_j {\cal D}[n_j]\rho$.
Here, $n_j=d^{\dagger}_jd_j$,
${\cal D}[n_j]\rho=n_j\rho n_j-\frac{1}{2}\{ n_j n_j, \rho\}$,
and the backaction-induced dephasing rate
\beqn\label{deph-rate}
\Gamma_j = 2\pi g_L g_R \left|\Del\Ome_{j}\right|^2
    e V_j \coth(\bet e V_j/2) .
\eeqn
At zero temperature and introducing the tunneling coefficients
${\cal T}_j=4\pi^2 g_L g_R |\Omega_j|^2$ and
${\cal T'}_j=4\pi^2 g_L g_R |\Omega'_j|^2$
(here we denote $\Omega'_j=\Omega_j+\Delta \Omega_j$),
we can reexpress the dephasing rate as
$\Gamma_j=(\sqrt{{\cal T}_j}-\sqrt{{\cal T'}_j})^2 eV_j/2\pi$.
In the latter feasibility estimates, we will use this compact expression.

To reveal the teleportation behavior,
we investigate the steady-state correlation function
\beq
S(t)=\left\langle M_2(t)M_1(0)\right\rangle_{ss},
\eeq
where the measurement operators are designed as
$M_1=n_1(1-n_2)$ and $M_2=n_2(1-n_1)$.
The meaning of $S(t)$ is clear.
In steady state, at some chosen {\it initial} moment ($t=0$),
if we find the left (right) dot occupied (unoccupied),
$S(t)$ predicts the probability
of finding the reversed occupation at time $t$,
say, the left dot empty and the right dot occupied.
This simply indicates an electron transfer (``teleportation")
between the {\it distant} dots, separated by the
{\it long} nanowire that supports the MBSs in the limit $\epsilon_M\to 0$.

Applying the master equation approach, we can easily
calculate the steady-state correlator $S(t)$.
Restricted in the odd-parity subspace,
one may first solve the master equation
to obtain the steady state $\rho_{ss}$.
Then, starting with $\rho_{ss}$ and after the ``$M_1$" measurement,
we propagate the resultant ``initial" state $\ket{100}\bra{100}$.
At the moment $t$, the probability of finding the state $\ket{001}\bra{001}$
is right the $S(t)$, which is given by $S(t)={\rm Tr}[M_2 \rho(t)]$.
In Fig.\ 2 we plot the result of $S(t)$ in both
the time domain and frequency space.
The damping oscillations in Fig.\ 2(a) show the
teleportation dynamics between the quantum dots,
under the influence of measurement backaction.
We remind again that this result is obtained in the limit of
$\epsilon_M\to 0$ which corresponds to a {\it long} nanowire.
So any change of $S(t)$ on short time scales
indicates a {\it teleportationlike} behavior.
Notice also that the maximal height of the peaks, less than unity,
is limited by the probability of finding ``$\ket{100}\bra{100}$"
in the steady state $\rho_{ss}$.
In Fig.\ 2(b) the same result in frequency space,
i.e., the Fourier transform of $S(t)$, is shown.
We notice a prominent ``dip" at $\omega=2\lambda$
and a relatively small ``peak" at $\omega=4\lambda$,
which are originated, respectively,
from the two harmonics involved in $S(t)$.
A simple way to understand this is from the Fourier spectrum of
$P^{(1)}_2(t)=\sin^4(\lam t)$,
which contains two harmonics with the above frequencies.

\begin{figure}[!htbp]
  \centering
  \includegraphics[width=5.5cm]{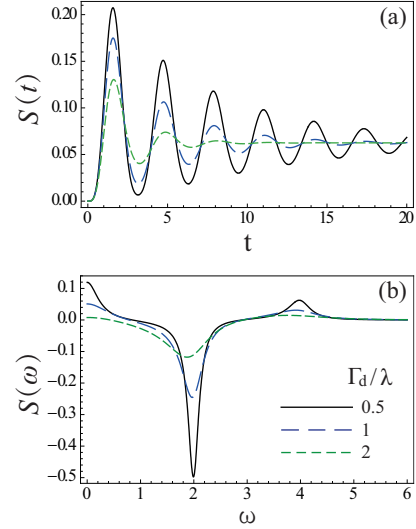}
  \caption{
Stationary cross-correlation in (a) the time domain
and (b) the frequency space.
We assume a symmetric setup and denote the
depahsing rate of \Eq{deph-rate} by $\Gamma_d$.
The units of frequency and time are, respectively,
$\lambda$ and $\lambda^{-1}$.  }\label{FIG2}
\end{figure}

\begin{figure}[!htbp]
  \centering
  \includegraphics[width=6.5cm]{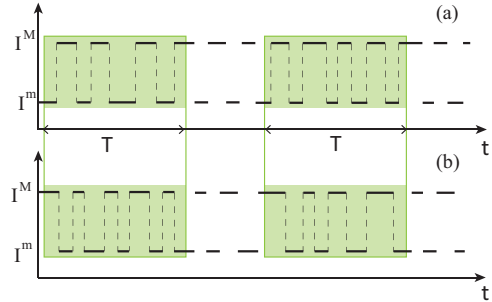}
  \caption{
Schematic output-currents in a strong-response regime.
Shaded are two examples of electron transfer
from the state $\ket{100}$ to $\ket{001}$.
We consider a symmetric setup which allows a
common maximum (minimum) current of $I^M$ ($I^m$). }\label{FIG3}
\end{figure}

The above dot-occupation correlator, $S(t)$,
is closely related to the following cross correlation
of detector currents,
$S_I(t)=\la M_{I,2}(t) M_{I,1}(0)\ra_{ss}$,
where $M_{I,1}=(I^M_1-I_1)(I_2-I^m_2)$
and $M_{I,2}=(I_1-I^m_1)(I^M_2-I_2)$.
$I^M_j$ is the (maximum) current through the $j$th point-contact
when the nearby dot is empty,
while $I^m_j$ is the (minimum) current when the dot is occupied.
Based on this type of correlation-function measurement,
in a relatively weak or intermediate response regime,
i.e., the detector does not {\it clearly} distinguish
the occupation/non-occupation of the nearby quantum dot,
one can obtain the result as shown in Fig.\ 2.
This corresponds to the continuous weak measurement,
which was also extensively studied in the
measurement of solid-state qubit \cite{Li0405,Sch01}.
In particular, the present two-detector coincident measurement
can be regarded as a counterpart of the
cross-correlation measurement of a solid-state qubit
by using two QPC detectors \cite{But05}.

On the other hand,
in a strong-response regime, the detector can definitely
distinguish whether the nearby quantum dot is occupied or not.
In this case, the output currents appear as the
telegraphic signals, as schematically shown by Fig.\ 3,
where we assumed a symmetric setup so that the maximum (minimum)
current is commonly denoted by $I^M$ ($I^m$).
Holding this type of data record, one has actually {\it unraveled}
the ensemble-averaged result.
In doing this, we should pay particular attention
to the process from $\ket{100}$ to $\ket{001}$,
passing through an intermediate state $\ket{010}$
in a short time interval $T$.
Or, more strikingly, as $T\to 0$ there may exist
a sudden ``jump" from $\ket{100}$ to $\ket{001}$.
These events, very clearly, reveal the teleportation phenomenon.


The above teleportation is an {\it unusual} consequence
of the quantum nonlocality of Majorana fermion.
Moreover, the {\it teleportation} process is
seemingly indicating a {\it superluminal} phenomenon.
In Ref.\ \cite{LeeDH08}, in order to rule out such possibility,
it was argued that, since a {\it classical exchange of information}
(the result of the coincident measurement) is necessary,
there is no superluminal transfer of information
in the observation of the teleportation effect.
However, we may notice here that, in the above joint measurements,
there is no need to perform the classical exchange of information
to confirm the result being $\ket{001}$ but not $\ket{111}$.
That is, we first keep the data record of the joint measurements
(as shown in Fig.\ 3). Then, we check whether there exists
such process that switches directly from $\ket{100}$ to $\ket{001}$,
but not through an intermediate state $\ket{111}$.
Since through our above analysis we do expect to find
such result (event) from the data record, and obviously
that event is {\it objective} despite that we did not confirm it
at {\it that moment} via any classical communication,
now the interesting problem is how we should interpret this result?
Unlike the argument in Ref.\ \cite{LeeDH08},
we would like to provide a different understanding.
Since the Majorana fermion is a quasiparticle excitation
in the presence of other electrons (background condensate
of electrons), we cannot conclude that the electron
appeared in the right quantum dot is the one
initially in the left dot.
This excludes the possibility of superluminal
electron transfer between the two remote quantum dots.
Therefore, the above result only demonstrates
the Majorana's nonlocality nature,
but does not imply a superluminal phenomenon.

{\it Feasibility.}---
The semiconductor InSb nanowire that may support the MBSs
has been utilized in the recent experiments \cite{Kou12}.
The InSb nanowire has a large $g$-factor (with $g\simeq 50$), and a strong
Rashba-type spin-orbit interaction (with energy $\sim$50 $\mu$eV).
Under proper magnetic field (e.g., $0.15$ Tesla), the Zeeman
splitting starts to exceed the {\it induced} superconducting
gap $\Delta\simeq 200~\mu{\rm eV}$ and thus to support the
emergence of MBSs at the ends of the nanowire.
Also, a low temperature such as $T=100$ mK can suppress the
thermal excitation of the Majorana zero mode to higher energy states.
If we tune the dot-MBS coupling ($\lambda$) to,
for instance, $20~\mu{\rm eV}$, the following estimates
show that the coherent oscillations of charge transfer
between the remote dots can be observed in the proposed setup.
Based on \Eq{deph-rate} and assuming a symmetric setup,
the measurement-backaction-induced dephasing rate reads
$\Gamma_d=(\sqrt{{\cal T}}-\sqrt{{\cal T'}})^2 V_d/2\pi$,
and the measurement ``signal" is simply given by
$\Delta I=I^M-I^m=({\cal T}-{\cal T'})V_d/2\pi$.
Here ${\cal T}$ (${\cal T'}$) denotes the transmission
coefficient through the point-contact barrier,
corresponding to the nearby dot being empty (occupied).
If we assume ${\cal T}=0.16$, ${\cal T'}=0.09$ and $V_d=1~{\rm mV}$,
a simple estimate gives $\Gamma_d \simeq 10~\mu{\rm eV}$
and $\Delta I \simeq 2.4~{\rm nA}$.
These results, reasonably, favor an implementation of the proposed
measurement scheme.

In summary, we analyzed a concrete realization and measurement
scheme to demonstrate the {\it teleportationlike} electron
transfer phenomenon mediated by Majorana fermion,
as a remarkable consequence of its inherently nonlocal nature.
Since all the major aspects of the proposed scheme are seemingly
within the reach of the state-of-the-art experiments in nowadays
laboratories, we expect that this striking phenomenon
can be demonstrated experimentally in the forthcoming future.

\vspace{0.5cm}
{\it Acknowledgments.}---
This work was supported by the NNSF of China (No. 101202101),
and the Major State Basic Research Project of China
(Nos.\ 2011CB808502 \& 2012CB932704).

%


\begin{references}


\bibitem{Kit01}   
A. Y. Kitaev, Physics-Uspekhi {\bf 44}, 131 (2001).
\bibitem{Lut10}   
R. M. Lutchyn, J. D. Sau, and S. Das Sarma,
Phys. Rev. Lett. {\bf 105}, 077001 (2010).
\bibitem{Ore10} 
Y. Oreg, G. Refael, and F. von Oppen,
Phys. Rev. Lett. {\bf 105}, 177002 (2010).

\bibitem{Zhang}
C. Zhang, S. Tewari, R. M. Lutchyn, and S. Das Sarma,
Phys. Rev. Lett. {\bf 101}, 160401 (2008).

\bibitem{Sau10}  
J. D. Sau, R. M. Lutchyn, S. Tewari, and S. Das Sarma,
Phys. Rev. Lett. {\bf 104}, 040502 (2010).
\bibitem{Sau12}  
J. D. Sau, S. Tewari, and S. Das Sarma,
Phys. Rev. B {\bf 85}, 064512 (2012).
\bibitem{Fu08}  
L. Fu and C. L. Kane, Phys. Rev. Lett. {\bf 100}, 096407 (2008).
\bibitem{Ali10}   
J. Alicea, Phys. Rev. B {\bf 81}, 125318 (2010).




\bibitem{Moore1991}
G. Moore and N. Read, Nuclear Physics B {\bf 360}, 362(1991).
\bibitem{Read2000}
N. Read and D. Green, Phys. Rev. B {\bf 61}, 10267 (2000).

\bibitem{DS01}  
K. Sengupta, I. Zutic, H. J. Kwon, V. M. Yakovenko, and S. Das Sarma,
Phys. Rev. B {\bf 63}, 144531 (2001).
\bibitem{Liu11}
D. E. Liu and H. U. Baranger, Phys. Rev. B {\bf 84},201308 (2011).
\bibitem{Bol07}   
C.J. Bolech and E. Demler,
Phys. Rev. Lett. {\bf 98}, 237002 (2007).
\bibitem{Law09}
K. T. Law, P. A. Lee, and T. K. Ng,
Phys. Rev. Lett. {\bf 103}, 237001 (2009).

\bibitem{LeeDH08}
S. Tewari, C. Zhang, S. Das Sarma, C. Nayak, and D. H. Lee,
Phys. Rev. Lett. {\bf 100}, 027001 (2008).


\bibitem{Nil08}
J. Nilsson, A. R. Akhmerov, and C. W. J. Beenakker,
Phys. Rev. Lett. {\bf 101},120403 (2008).

\bibitem{Fu09}
L. Fu and C. L. Kane, Phys. Rev. B {\bf 79}, 161408 (2009).


\bibitem{Kou12}
V. Mourik {\it et. al}, Science {\bf 336}, 1003 (2012).

\bibitem{MF2}
M. T. Deng, C. L. Yu, G. Y. Huang, M. Larsson, P. Caroff,
and H. Q. Xu, arXiv:1204.4130.
\bibitem{MF3}
L. P. Rokhinson, X. Liu, and J. K. Furdyna, arXiv:1204.4212.
\bibitem{MF4}
A. Das, Y. Ronen, Y. Most, Y. Oreg, M. Heiblum,
and H. Shtrikman, arXiv:1205.7073.



\bibitem{Ens06}
S. Gustavsson {\it et. al}, Phys. Rev. Lett. {\bf 96}, 076605 (2006);
T. Fujisawa {\it et. al}, Science {\bf 312}, 1634 (2006).

\bibitem{Li0405}
X. Q. Li, W. K. Zhang, P. Cui, J. S. Shao,
Z. S. Ma, and Y. J. Yan, Phys. Rev. B {\bf 69}, 085315 (2004);
X. Q. Li, P. Cui, and Y. J. Yan,
Phys. Rev. Lett. {\bf 94}, 066803 (2005).
\bibitem{Sch01}
Yu. Makhlin, G. Sch\"on, and A. Shnirman,
Rev. Mod. Phys. {\bf 73}, 357 (2001).
\bibitem{But05}
A. N. Jordan and M. B\"uttiker, Phys. Rev. Lett. {\bf 95}, 220401 (2005).

\end{references}
\end{document}